\documentclass[10pt,leqno]{amsart}
\usepackage{graphicx}
\usepackage{indentfirst,csquotes}
\usepackage[foot]{amsaddr}

\usepackage{amssymb,amsthm,amsmath}
\usepackage{xcolor,paralist,hyperref,titlesec,fancyhdr,etoolbox}

\titleformat{\section}[display]{\normalfont\huge\bfseries\centering}{\centering\chaptertitlename\thechapter}{10pt}{\Large}
\titlespacing*{\section}{0pt}{0ex}{0ex}

\hypersetup{ colorlinks=true, linkcolor=black, filecolor=black, urlcolor=black }

\RequirePackage[numbers,sort&compress]{natbib}
\newlength{\biblabelwidth}
\RequirePackage{jabbrv}

\begin{document}
\bibliographystyle{osajnl}

\title[Mid-infrared SNSPDs]{Low-noise single-photon counting superconducting nanowire detectors at infrared wavelengths up to 29 \textmu m} 
\author[]{Gregor G. Taylor\textsuperscript{1}}
\author[]{Alexander B. Walter\textsuperscript{1}}
\author[]{Boris Korzh\textsuperscript{1}}
\author[]{Bruce Bumble\textsuperscript{1}}
\author[]{Sahil R. Patel\textsuperscript{1}}
\author[]{Jason P. Allmaras\textsuperscript{1}}
\author[]{Andrew D. Beyer\textsuperscript{1}}
\author[]{Roger O'Brient\textsuperscript{1}}
\author[]{Matthew D. Shaw\textsuperscript{1}}
\author[]{Emma E. Wollman\textsuperscript{1}}
\date{\today}
\address{\textsuperscript{1}Jet Propulsion Laboratory, California Institute of Technology, 4800 Oak Grove Dr., Pasadena, CA}
\maketitle

\let\thefootnote\relax
\footnotetext{\textit{Email}: gregor.g.taylor@jpl.nasa.gov} 

\begin{abstract}
We report on the extension of the spectral sensitivity of superconducting nanowire single-photon detectors to a wavelength of 29 \textmu m. This represents the first demonstration of a time correlated single-photon counting detector at these long infrared wavelengths. We achieve saturated internal detection efficiency from 10 to 29 \textmu m, whilst maintaining dark count rates below 0.1 counts per second. Extension of superconducting nanowire single-photon detectors to this spectral range provides low noise and high timing resolution photon counting detection, effectively providing a new class of single-photon sensitive detector for these wavelengths. These detectors are important for applications such as exoplanet spectroscopy, infrared astrophysics, physical chemistry, remote sensing and direct dark-matter detection.
\end{abstract} 

\bigskip
\medskip
\noindent 
Single-photon counting detectors (SPDs) represent the ultimate sensitivity in photodetector technology at a given wavelength. They have become central to multiple fields of science from quantum information processing and quantum communications to fundamental physics~\cite{monroe2002quantum,Sasaki:11,PhysRevLett.115.250401}. The detection energy threshold of a detector governs the maximum wavelength of an absorbed photon that will result in a detection being registered. Lowering this energy threshold to count longer wavelength photons has been proposed as a potential benefit to multiple applications such as quantum sensing~\cite{dello2022advances}, light detection and ranging (LIDAR)~\cite{Taylor2019}, quantum communications~\cite{aellen2008feasibility}, physical chemistry~\cite{Chen2017} and astrophysics~\cite{wollman2021recent}. In particular, the fields of infrared astronomy and dark matter detection are driven by the ability to perform low-noise single-photon counting at the longest possible wavelength. However, even at the shorter end of the mid-infrared spectral region (2 to 10~\textmu m), there is a shortage of photon counting SPD options, with HgCdTe avalanche photodiodes that operate in single-photon counting mode\cite{anderson2022recent} and superconducting nanowire single-photon detectors (SNSPDs~\cite{gol2001picosecond, Verma2021, morozov2022superconducting}) the only real technologies available. Looking to longer wavelengths, to the farther end of the mid-infrared (10 to 25~\textmu m) and into the far-infrared (above 25~\textmu m), there exists no demonstrated technology for single-photon counting. In this work we are primarily concerned with photon-counting detector development in this 10 to 30~\textmu m spectral region, which will be referred to hereafter as the mid-infrared. 

Detectors capable of single-photon detection at wavelengths at this longer end of the mid-infrared spectrum, if available, have potential to become revolutionary for a variety of applications, including astronomy, dark matter searches and physical chemistry. In  exoplanet spectroscopy~\cite{seager2008exoplanet}, transiting exoplanets can be detected with far better contrast in the mid-infrared than in the visible~\cite{boccaletti2005imaging}, and their atmospheric composition can be deduced~\cite{kaltenegger2019characterize}. Working at these longer wavelengths also allows less stringent requirements on the optics and control of a characterization system~\cite{martin2012high}. Proposals for characterization of exoplanets in the mid-infrared have recently been studied and concluded that a detector with high signal-to-noise ratio and a long wavelength cut-off of at least 18.5~\textmu m would be optimal~\cite{konrad2021large}, with some suggesting 25~\textmu m as an even better choice~\cite{quanz2018exoplanet}. Photon-counting detectors are desirable for this application, because their digital-like signals eliminate several sources of instability in analog detectors that can make it difficult to detect the small changes in the host star’s spectrum over the time of a transit~\cite{wollman2021recent}. In addition to the above, a wide variety of atoms, ions and molecules present in astronomical objects have emissions in this spectral range~\cite{peeters2002iso, palotas2020infrared}. Beyond astronomy applications, direct dark-matter (DM) detection experiments would benefit from the sensitive single-photon detection that detectors operating in this spectral range would offer~\cite{golwala2022novel}. Recent experimental searches for DM candidates such as axions and dark photons like the BREAD~\cite{PhysRevLett.128.131801} and LAMPOST~\cite{chiles2022new} experiments would benefit from detectors such as SNSPDs. Lowering the detection energy threshold of these detectors whilst maintaining low dark count rates has been shown to be a promising path forward towards probing new parameter space for DM candidates~\cite{hochberg2022new}.
Lastly, in physical chemistry, low noise-equivalent power single-photon detectors operating at mid-infrared wavelengths would allow probing of the whole range of vibrational frequencies found in the molecular finger print region with single-photon sensitivity~\cite{lau2023superconducting}.

To date, however, even non-photon counting photodetectors in the 10 to 30~\textmu m wavelength range have proved difficult to attain. Blocked impurity band (BIB) detector arrays are common in astronomy applications and have been utilized out to 28~\textmu m with Si:As~\cite{ressler2008performance} and demonstrated out to 38~\textmu m with Si:Sb~\cite{wada2010mid}. They are, however, susceptible to a variety of issues -- such as reset anomalies, last-frame effects, droop and drift -- that lead to instability in the count rate~\cite{ressler2015mid,greene2016characterizing} and are inherently limited by readout noise. HgCdTe photovoltaic detectors have been demonstrated at 15~\textmu m~\cite{cabrera2019characterization} and have been proposed out to 20~\textmu m and beyond. However, as the Hg content of such detectors increases to lower the bandgap, they become increasingly soft and prone to defects~\cite{pipher2021mid}. III-V material type-II superlattice detectors could provide a pathway to a more robust detector but thus far have not been realized~\cite{rogalski2019type}. Alternative superconducting detector options such as kinetic inductance detectors (KIDs) are also under development to operate in the mid-infrared~\cite{roellig2020mid} but are yet to be experimentally demonstrated at these wavelengths~\cite{hailey2021kinetic} and are currently limited by two-level system noise~\cite{perido2020extending}. Transition edge sensors (TESs) are sensitive in the mid-infrared but have not been demonstrated in photon-counting mode and require complex SQUID-based readout~\cite{lamas2013nanosecond}. It is worth noting that due to all of the above technologies not yet being true single-photon counting detectors at mid-infrared wavelengths, they can not reach the fundamental noise-equivalent power limits and are unsuitable for experiments such as dark matter searches, where single-photon counting is required.

SNSPDs are currently the gold standard for near-infrared photon-counting in terms of system detection efficiency~\cite{Reddy2020, hu2020detecting, chang2021detecting}, timing jitter~\cite{korzh2020demonstration}, dark count rate~\cite{chiles2020superconducting, wang2022twin} and maximum count rate~\cite{craiciu2023high}. A characteristic of a high performance SNSPD is the presence of a saturated plateau in the detection efficiency against bias measurement, indicating that every photon absorbed in the nanowire results in an output pulse being produced when the detector is biased in this saturated regime. When operated in this region, the efficiency is somewhat immune to bias current fluctuations, making SNSPDs highly stable devices~\cite{wollman2021recent}. In addition, they exhibit zero readout noise and are truly single-photon sensitive, time-resolved photon counting devices.  

SNSPDs have been developed in recent years as a promising technology for the mid-infrared wavelength range as the device properties can be tuned to allow detection of single low-energy photons. Although potential long-wavelength photon detection with SNSPDs has been predicted for many years~\cite{Marsili2012a}, it still remains an engineering challenge to realize detectors with high detection efficiency due to the required fabrication tolerances and necessary material development. To date, unity internal detection efficiency has been demonstrated out to wavelengths of 10~\textmu m~\cite{Verma2021} in short nanobridges and 7.4~\textmu m in large active-area meander structures~\cite{colangelo2022large}. Other research groups have also demonstrated sensitivity at mid-infrared wavelengths up to 10~\textmu m with unity internal efficiency at shorter wavelengths~\cite{korneev2012nbn, pan2022mid, chen2020mid}. SNSPDs with high system detection efficiencies (system detection efficiencies include optical losses) have also been reported at the shorter end of the mid-infrared spectrum~\cite{chang2022efficient}.

In this work, we optimize the film stoichiometry, geometric parameters and operating temperature to demonstrate unity internal detection efficiency at wavelengths up to 29~\textmu m ($E_{\mathrm{photon}}$ = 43~meV, $\nu$ = 10.3~THz) in SNSPDs, for the first time.

\bigskip
\section*{Results}
\bigskip
\subsection*{Material Engineering}
The detection energy threshold of an SNSPD scales with the characteristic energy, $E_{0} = 4N(0)(k_{b}T_\mathrm{c})^{2}V_{0}$, where $N(0)$ is the density of states per spin at the Fermi level in the normal state, given by $N(0) = (2\rho e^{2}D)^{-1}$, $k_{b}$ is Boltzmann's constant, $T_\mathrm{c}$ is the critical temperature, $V_{0}$ is the characteristic volume, $D$ is the diffusion coefficient, $\rho$ is the resistivity of the film and $e$ is the electron charge~\cite{vodolazov2017single}. By increasing the Si content of the WSi superconducting film we can increase the resistivity of the film (lower the free carrier density), thus reducing the density of states, $N(0)$. In the detection process, this results in the deposited photon energy being divided amongst a smaller number of quasiparticles within a given volume, increasing the quasiparticle energy and facilitating the suppression of superconductivity~\cite{Verma2021}. Increasing the Si content also reduces the $T_\mathrm{c}$ of the material. Additionally, we can utilize thinner films, which both further lowers the $T_\mathrm{c}$  and also the characteristic volume $V_{0}$. This volume factor is a complicated function of several device parameters including bias current, electron-phonon and phonon escape timescales and wire geometry, but as the film thickness is the same order as the coherence length, the system can be treated as quasi-two-dimensional such that $V_{0}$ scales linearly with thickness. Combining the above approaches, we see we can effectively lower the characteristic energy in a three-fold fashion by minimizing $N(0)$, $T_\mathrm{c}$ and $V_{0}$.

We deposited 3~nm WSi thin films on Si substrates with 240~nm of thermal oxide via sputtering with a 30:70 W:Si compound target at a power of 130~W and a pressure of 5 mTorr. A 20 nm a-Si cap was added as a passivation layer. The resulting film had sheet resistance, $R_{s}$ = 1.16~k$\Omega$, RRR (R$_{300K}$/R$_{20K}$) of 0.98 and $T_\mathrm{c}$ = 1.3~K. We take $D$ to be 0.53~cm$^{2}$s$^{-1}$ as in \cite{colangelo2022large} for the same target. This gives us a $N(0)$ of 17.3~eV$^{-1}$nm$^{-3}$ and a $4N(0)(k_{b}T_\mathrm{c})^{2}$ value of 0.81~\textmu eV nm$^{-3}$. Comparing this to Ref.~\cite{colangelo2022large} we see that we have reduced the $N(0)$ very slightly and, due to the lowering of the $T_\mathrm{c}$, have substantially lowered the $4N(0)(k_{b}T_\mathrm{c})^{2}$ value. A reduction by a factor of 2.9 is achieved compared to the work in~\cite{colangelo2022large} and a factor of 4.9 reduction when compared to~\cite{Verma2021}. This should result in improved detection of low-energy photons. 

The film was patterned into nanowires of 80 nm width via electron-beam lithography with ZEP530A resist and a reactive-ion etch~\cite{colangelo2022large}. The wire width was chosen as a good trade-off between keeping the cross-section of the nanowire small for efficient photon detection, whilst minimizing constrictions caused by fabrication imperfections, which are more prevalent at narrower wire widths~\cite{Frasca2019}. Meander structures, where the nanowire was meandered back and forth in order to provide a larger active area, and straight nanobridges of smaller active area were fabricated. The pitch of the meanders was 240~nm, and they covered an active area of 11~\textmu m by 10~\textmu m. The nanobridges were 20~\textmu m long and included a separate wide meander section to add inductance and prevent latching~\cite{annunziata2010reset}.

\subsection*{Photoresponse}
\begin{figure}
        \centering
        \includegraphics[width = \columnwidth]{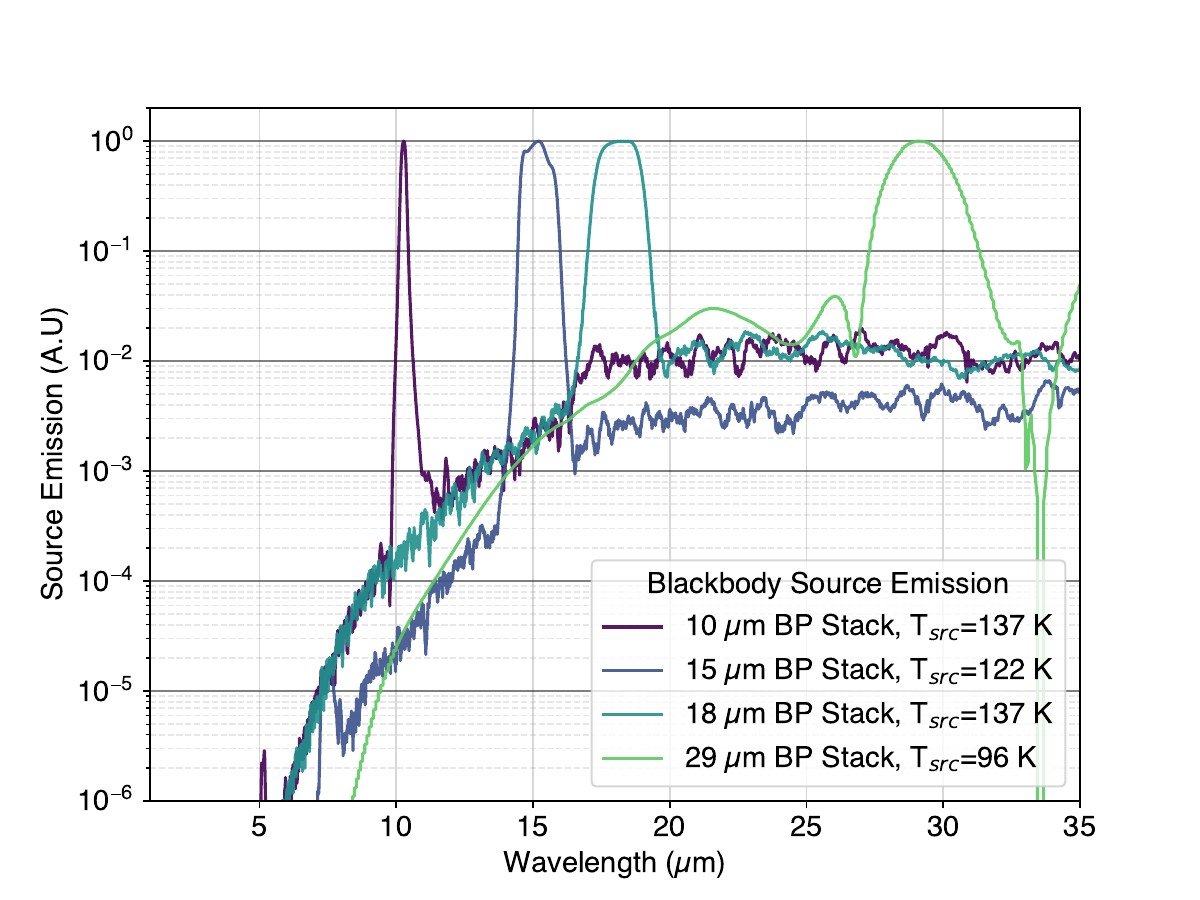}
        \caption{Blackbody source emission with various filter stacks for different target wavelengths. The temperature of the source is noted in the legend. The data here has been smoothed with a Savitzky–Golay filter for clarity; see the Supplementary Information for the raw data.}
        \label{fig:filterTX}
\end{figure}

The devices were cooled to 250 mK with a $^{3}$He sorption cryocooler. A pulsed cryogenic blackbody source, operated at 1 Hz with a duty cycle of 37.5\% (Axertis EMIRS50, see Supplementary Information and Ref.~\cite{Walter2023} for more information), was mounted on the 4~K stage and flood illuminated the device through a filter stack. For each wavelength except 29~\textmu m, a bandpass filter at the wavelength of interest was combined with ZnSe short pass filters to exclude long wavelength photons. For the 29~\textmu m measurements, two 29~\textmu m bandpass filters were combined. The resulting filter stacks were characterized in a Fourier-transform infrared (FTIR) spectrometer to ensure unwanted short-wavelength photons were effectively suppressed. We calculated the temperature of the blackbody source by calibrating the temperature vs. resistance curve and monitoring the resistance during operation. Fig.~\ref{fig:filterTX} shows the calculated blackbody emission spectra through the various filter stacks. More details on the blackbody source and characterization of filters is shown in the Supplementary Information.
Devices were biased via a cryogenic bias tee and output pulses were read out via a cryogenic amplifier (Cosmic Microwave CMTLF1) at 4 K and a room temperature low-noise amplifier (Mini-Circuits CKL-1R5+). 

\begin{figure}
        \centering
        \includegraphics[width = \columnwidth]{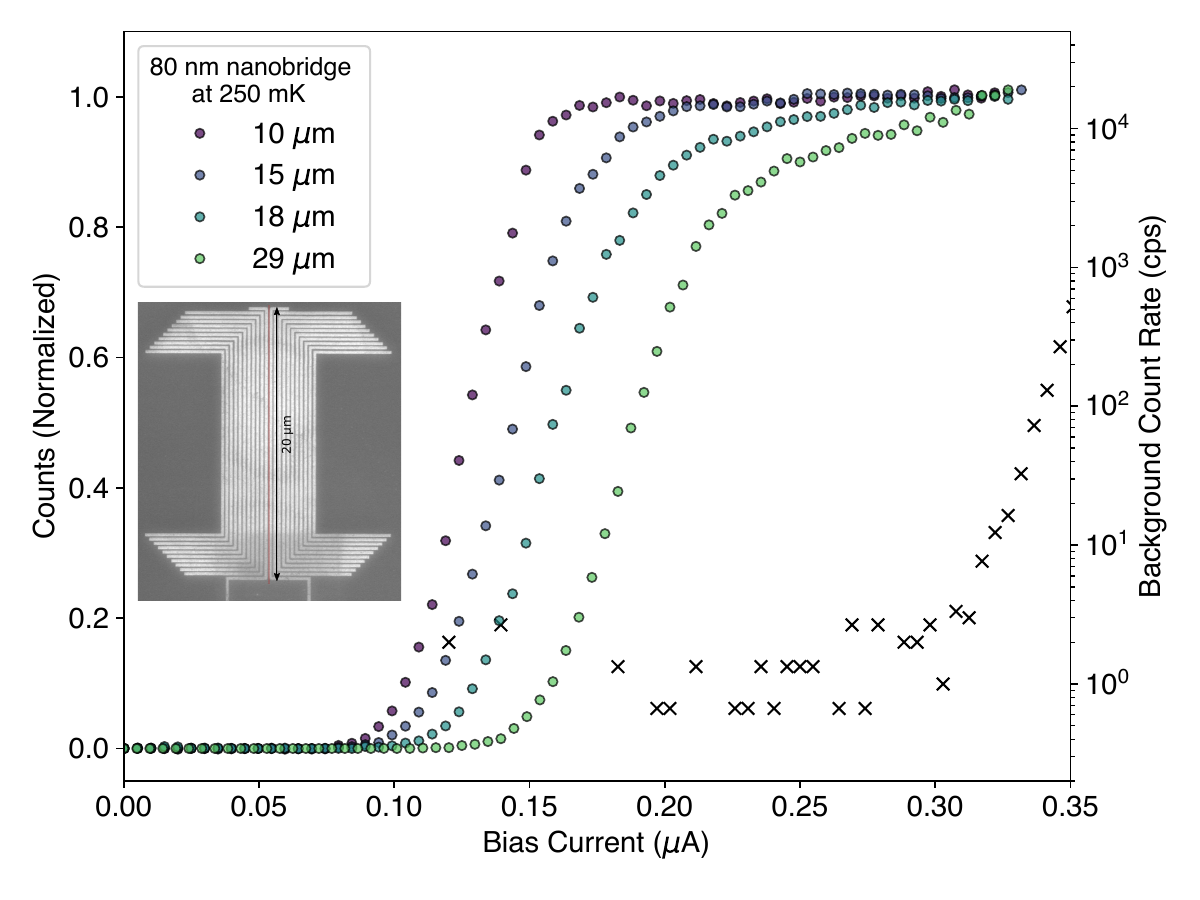}
        \caption{Normalized PCR curves for 80 nm wide nanobridge at 250 mK for wavelengths of 10 \textmu m, 15~\textmu m, 18~\textmu m and 29~\textmu m. The saturated count rates were 6 kcps, 12 kcps, 7 kcps and 8 kcps for each wavelength respectively. BCR is shown in black $\times$'s on the right y-axis. The inset shows an SEM of the device with the active area highlighted in red. The remaining structures visible are for proximity error correction during the electron beam lithography.}
        \label{fig:nanbridgePCR}
\end{figure}

Photon count rate (PCR) vs. bias current curves were taken for the 80~nm nanobridge for wavelengths of 10~\textmu m, 15~\textmu m, 18~\textmu m and 29~\textmu m and the results are shown in Fig.~\ref{fig:nanbridgePCR}. The blackbody source was pulsed (1 Hz, 37.5\% duty cycle) so that the background count rate (BCR) could be measured in the same period as the PCR during the dark portions of the sources cycle. This minimized any temperature variation between measurements, as well as the heat load on the 4~K stage. For each data point the BCR was subtracted from the counts measured to give the PCR. The BCR is also shown in Fig.~\ref{fig:nanbridgePCR}. 

Saturation of the PCR curve as the bias current is increased is observed for all wavelengths in the nanobridge, indicating unity internal detection efficiency (IDE). A slight slope is observed in the longer wavelength measurements, which is likely due to the absorption of longer wavelength photons beyond 29 \textmu m due to imperfect optical filtering, as well as absorption of photons in the tapered region at the ends of the nanobridges.   

\begin{figure}
        \centering
        \includegraphics[width = \columnwidth]{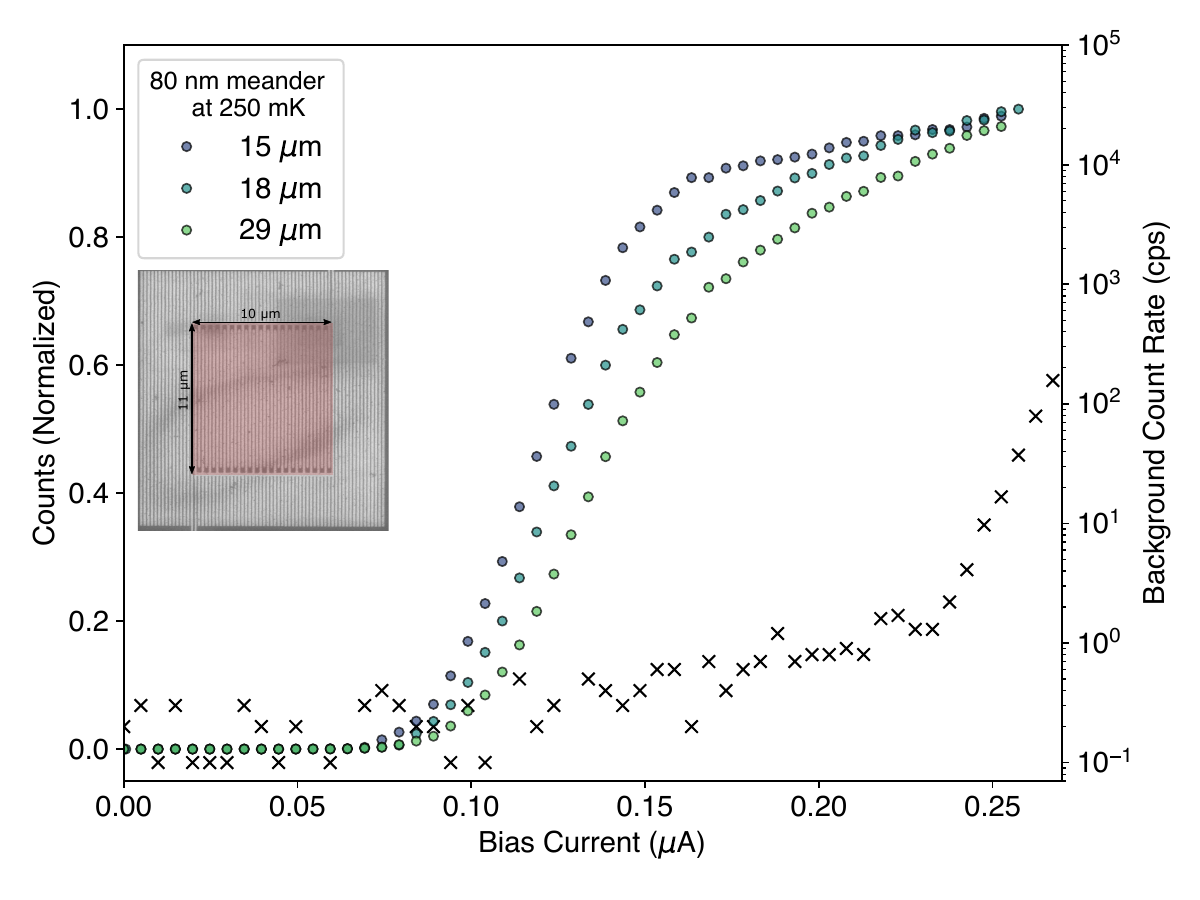}
        \caption{Normalized PCR curves for 80 nm wide 10$\times$11 \textmu m meander at 250 mK for wavelengths of 15~\textmu m, 18~\textmu m and 29~\textmu m. The saturated count rates were 23 kcps, 35 kcps and 34 kcps for each wavelength respectively. BCR is shown in black $\times$'s on the right y-axis. The inset shows an SEM of the device with the active area highlighted in red. The remaining structures visible are for proximity error correction during the electron beam lithography.}
        \label{fig:meanderPCR}
\end{figure}

The 80 nm meanders also exhibited photoresponse at 15~\textmu m, 18~\textmu m and 29~\textmu m, and the PCR curves are shown in Fig.~\ref{fig:meanderPCR}. The meander structure leads to increased chance of constrictions in the bends and along the length of the nanowire during fabrication and results in a reduced ability to saturate. Both the 15~\textmu m, 18~\textmu m and 29 \textmu m light, however, show an inflection point and plateau region, although the PCR continues to climb slightly after the plateau is reached. This is likely predominantly due to local regions of increased switching current (I$_{\mathrm{sw}}$)~\cite{doi:10.1063/1.4881981} caused by constrictions, with a contribution from long wavelength photons due to imperfect filtering also likely.

\subsection*{Dark Counts}
The dark counts of a detector can be broadly split into the intrinsic detector dark count rate (DCR) and the background count rate (BCR). The DCR is the counts that would be registered by the device in the absence of photon absorption and the BCR includes the counts registered due to external radiation coupling into the experimental environment. The BCR is a function of the optical coupling scheme used to direct signal to the detector, so it is highly dependent on the optical design of the detector system. It is useful to characterize both the BCR and the DCR. In initial testing, the samples were mounted onto a copper sample plate with no shielding from higher temperature stages and components. This resulted in a BCR on the order of 10$^{2}$ counts per second (cps) for the nanobridge and 10$^{3}$~cps for the meander, when biased on the saturated plateau (see Supplementary Information for data).

In order to lower the BCR, a new detector mounting package was designed and manufactured. The detector is housed in a gold-plated, oxygen-free high-conductivity (OFHC) copper box with a 1/2" aperture. The filter stack was then mounted directly onto the aperture. Care was taken that no dielectric materials were included inside the package due to transmission in the mid-infrared and the likelihood of Cherenkov-radiation induced dark counts~\cite{du2022sources}. The resulting BCR of the nanobridge and meander in the new package with the 18 µm filter stack in place is shown in Fig.~\ref{fig:nanbridgePCR} and Fig.~\ref{fig:meanderPCR} respectively. The BCR was integrated for 10 seconds per bias point. A reduction of the BCR to sub-10~cps was achieved on the saturated plateau for both detector geometries. More information on the packaging is available in the Supplementary Information.

\begin{figure}
        \centering
        \includegraphics[width = \columnwidth]{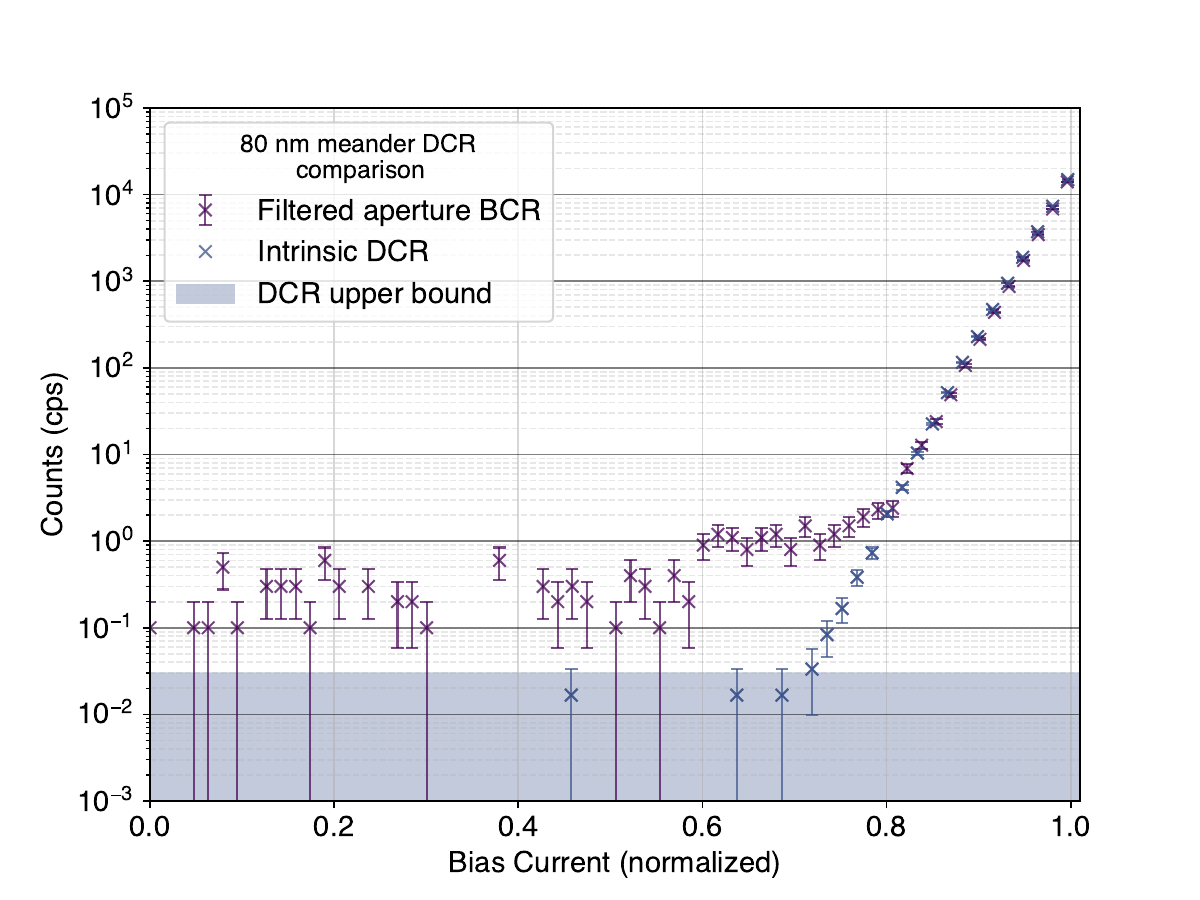}
        \caption{DCR comparison between the detector package with a filtered aperture and the closed box intrinsic DCR. Bias is normalized to the I$_{\mathrm{sw}}$ to account for slight DC offset between cooldowns. The upper bound for the DCR where no counts were observed in the integration period is shown in the shaded orange region.}
        \label{fig:DCRComparison}
\end{figure}

It is also useful to quantify the intrinsic DCR, that is the DCR in a completely sealed environment, in order to put a lower limit on the achievable BCR. For completely light-tight experiments such as dark matter detection, the intrinsic DCR sets a limit on the noise floor of the experiment~\cite{chiles2022new}. To do this, the meander device was mounted in a similar package with no aperture. To quantify the intrinsic DCR, a counts vs. bias current curve was taken with a 60 second integration time per bias point. The resulting data is shown in Fig.~\ref{fig:DCRComparison} alongside the data from a previous BCR measurement with the 18 \textmu m optical filter stack in place. The intrinsic DCR is measured to be below 0.1~cps just before the exponential increase as the current approaches I$_{\mathrm{sw}}$. An upper limit was calculated for the bias points where no counts were registered in the integration period using the modified Wald Barker method \cite{patil2012comparison} using a 68\% confidence interval. To further bound this value we used an oscilloscope (LeCroy, 6~GHz BW) to perform a long integration DCR measurement over a timescale of 2 hrs. The oscilloscope was used to digitize the output pulses so they could be examined to ensure actual detector clicks and not spurious noise were triggering the counter. A bias point of 0.22~\textmu A (0.73 on normalized Fig.~\ref{fig:DCRComparison}) was selected as this was on the saturated plateau for all wavelengths but before the exponential onset of dark counts - i.e a suitable operating bias point for this detector. Over the 2~hr integration time, 130 events were recorded, 9 of which were determined to be caused by noise/spurious signals on the readout lines. This gives a DCR of 1.6$\times$10$^{-2}$~cps, consistent with the data obtained in the 60 second integration measurement. 

\bigskip
\section*{Discussion}
\bigskip
The presence of a saturated plateau region in the PCR curve of an SNSPD indicates unity IDE - that is every photon absorbed by the nanowire results in a output pulse being produced. The nanobridges exhibited unity IDE for all wavelengths tested, whilst the meanders showed an inflection point and flattening of the PCR curve, indicating close-to unity IDE, but did not show a full saturated plateau. This was likely due to photon absorption events in the bends and fabrication imperfections along the length of the nanowire. Meander bends and fabrication imperfections can cause constrictions, resulting in local regions of increased current crowding. This leads to a suppression of the switching current at certain points in the wire, and hence a spatial variation to the bias point at which unity IDE is achieved. 

The total detection efficiency depends on both the IDE and the optical coupling efficiency, so the next step in the development of these detectors is to increase the absorption of incident radiation by the nanowire. SNSPDs are usually fabricated in an optical cavity~\cite{baek2009superconducting} to enhance their absorption efficiency, but this proves difficult in this wavelength range due to the unavailability of suitable dielectrics and thickness requirements for efficient coupling. Antenna coupling~\cite{heath2015nanoantenna} and microlenses~\cite{xu2021superconducting} could provide possible pathways to increasing the absorption efficiency at these long wavelengths. 

\begin{figure}[ht]
        \centering
        \includegraphics[width = \columnwidth]{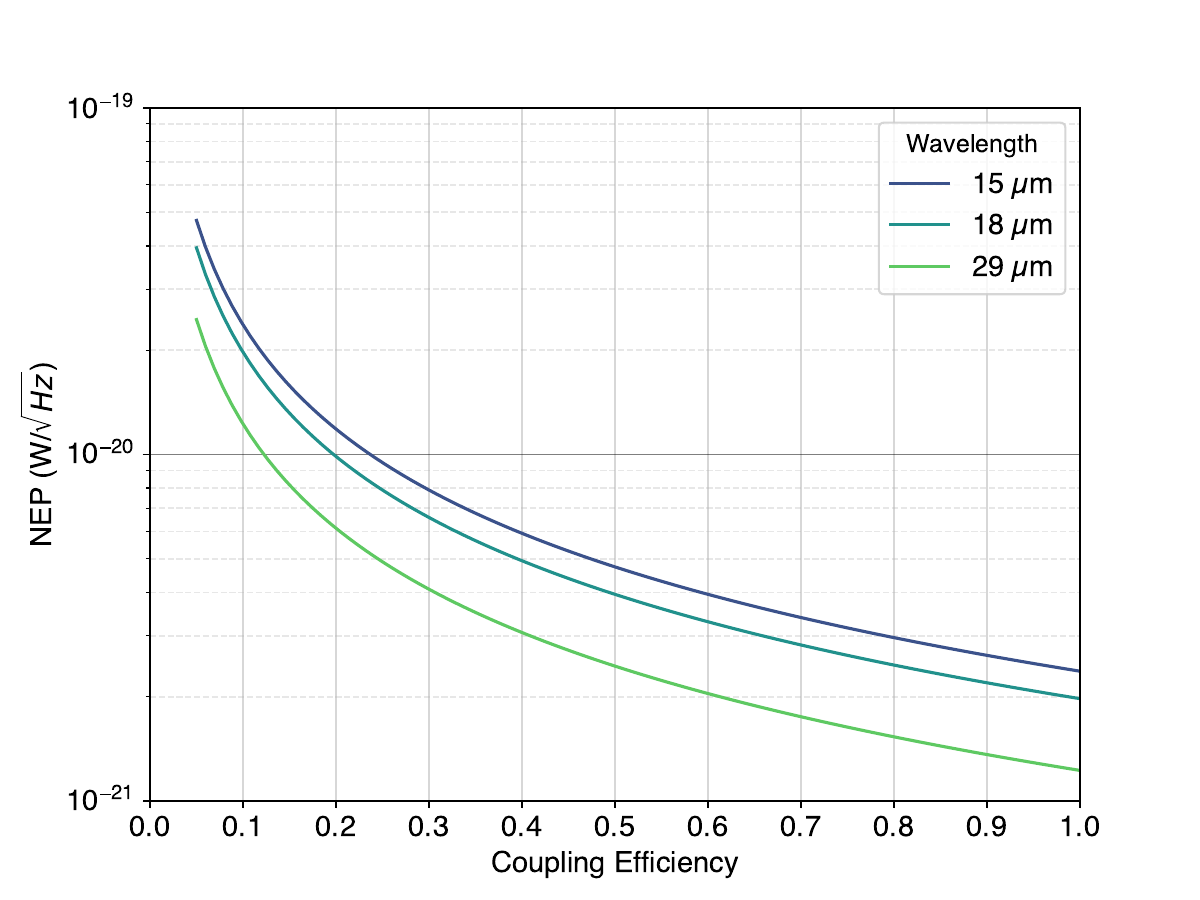}
        \caption{Calculation of noise-equivalent power for our detector at 15~\textmu m, 18~\textmu m and 29~\textmu m wavelength, given the measured DCR and internal detection efficiency saturation, for a range of optical coupling efficiencies.}
        \label{fig:NEPCalc}
\end{figure}

In the mid-infrared, photodetectors are typically characterized by their noise-equivalent power (NEP). The NEP for a single-photon detector is calculated by~\cite{hadfield2009single}:
\begin{equation}
    NEP=\frac{h\nu}{\eta}\sqrt{2D}
\end{equation}
where $\eta$ is the efficiency of the detector and $D$ is the DCR of the detector. As unity IDE was demonstrated in this work, $\eta$ is entirely dependent on the optical coupling efficiency. If we assume an achievable 50\% coupling efficiency and take the DCR measured during the 2 hr integration, we calculate an NEP of 2.5$\times$10$^{-21}~$W/$\sqrt{\mathrm{Hz}}$ on the saturated plateau at 29 \textmu m wavelength. To put this number in context, a proposed requirement to meet the science goals of a future flagship class infrared space telescope is an NEP of 3$\times$10$^{-20}~$~W/$\sqrt{\mathrm{Hz}}$~\cite{hailey2021kinetic}, and our detector would compare favourably with the best KIDs in the 1-10 THz range~\cite{baselmans2022ultra}. Fig.~\ref{fig:NEPCalc} shows the calculation of NEP for a variety of coupling efficiencies showing that even with a sub-10\% optical coupling, NEPs in the low 10$^{-20}$~W/$\sqrt{\mathrm{Hz}}$ range would be achievable.

Due to the low I$_\mathrm{sw}$ of these devices, a natural result of the low T$_\mathrm{c}$ and nanowire geometry, pulse heights after amplification were in the 10~s of mV scale. Whilst large enough to be satisfactorily counted, a larger signal-to-noise ratio (SNR) would be beneficial to separate the true detector clicks from spurious noise. Impedance matching tapers on the nanowire ends~\cite{zhu2019superconducting} and a superconducting nanowire avalanche photodetector (SNAP)~\cite{marsili2011single} architecture would be readily applicable techniques here to increase the output pulse height and improve the SNR. This would allow a higher triggering threshold which would further reduce the DCR, as shown earlier where the oscilloscope was used to veto noise counts.

The temperature dependence of the detectors was characterized by taking a different 80~nm wide nanobridge device from the wafer and cooling to 100~mK in an adiabatic demagnetization refrigerator (ADR). By ramping the magnetic field, the temperature could be tuned to observe the behavior of the detector. A PCR curve for an illumination wavelength of 10~\textmu m was taken for different temperatures and a summary of the results is shown in Table.~\ref{tab:tempDependenceTab}. The current where the detector began to detect photons (turn-on current) was 0.12~\textmu A for this device. The current where saturated behaviour began was 0.235~\textmu A and the plateau length was calculated by subtracting this from the I$_\mathrm{sw}$ for each temperature. The detector showed no degradation in performance between 100~mK and 306~mK. At 481~mK, the I$_\mathrm{sw}$ was slightly reduced but saturated performance was still observed with a long plateau indicating that 15~\textmu m, 18~\textmu m and 29~\textmu m  plateaus would likely still be achievable. At 780~mK, the detector performance had degraded beyond use with no plateau observed. From this is it reasonable to conservatively conclude, given the $T_\mathrm{c}$ of the device was 1.3~K, that these detectors will exhibit saturated performance in the mid-infrared at a fraction of T$_\mathrm{c}$ of around 0.4 and lower. Performance at higher temperatures may be feasible at a reduced internal detection efficiency. With this knowledge, it is likely that we can tune the stoichiometry of the WSi film by further increasing the Si content, which would result in films with even lower $T_\mathrm{c}$, and still operate them without requiring the use of a dilution refrigerator. This would further improve the low energy sensitivity, as outlined earlier in this paper.

\begin{table}[ht]
    \centering
    \begin{tabular}{c|c|c}
        \textbf{Temperature (mK)} & \textbf{I$_\mathrm{sw}$ (\textmu A)} & \textbf{Plateau Length (\textmu A)}\\
        100 & 0.6 & 0.37 \\
        306 & 0.59 & 0.36 \\
        481 & 0.48 & 0.25 \\
        780 & 0.2 & N/A \\
    \end{tabular}
    \medskip
    \caption{Temperature dependence on the I$_\mathrm{sw}$ and plateau presence for an 80 nm wide nanobridge at 10 \textmu m wavelength illumination.}
    \label{tab:tempDependenceTab}
\end{table}

In conclusion, we have demonstrated that SNSPDs can offer low-noise single photon counting operation in the mid-infrared. By demonstrating saturated internal detection efficiency at wavelengths as long as 29~\textmu m, whilst maintaining DCR as low as 1$\times$10$^{-2}$~cps, we have shown that they are capable of achieving excellent detection metrics in this wavelength range. Investigation of the temperature dependence suggests the techniques utilized in this work are readily extendable to even longer wavelengths. This work promises a new class of single-photon detectors for the mid-infrared spectral range with the inherent low timing jitter, high stability, low noise and potential for high count rates that SNSPDs can provide. 

\subsection*{Supplementary information}
Please see the Supplementary Information for full details on the thermal source, optical filtering and device packaging.
\subsection*{Acknowledgments}
This research was performed at the Jet Propulsion Laboratory, California Institute of Technology, under contract with the National Aeronautics and Space Administration (NASA - 80NM0018D0004). Support for this work was provided in part by the DARPA DSO Invisible Headlights program and the NASA ROSES-APRA program. A.B. Walter's research was supported in part by an appointment to the NASA Postdoctoral Program at the Jet Propulsion Laboratory, administered by Universities Space Research Association under contract with NASA. The authors would like to thank Lautaro Narv\'{a}ez for engineering assistance and Karl K. Berggren and Varun B. Verma for in depth technical discussions. G.G. Taylor would like to thank Tim Crawford for operation of the fourier-transform spectrometer for filter characterization and illuminating discussions. We also thank Andrew S. Mueller and Ioana Craiciu for their helpful comments on the manuscript.

\noindent \textcopyright 2023. California Institute of Technology. Government sponsorship acknowledged.
\\

\bibliography{sn-bibliography}

\begin{thebibliography}{10}

\bibitem{monroe2002quantum}
C.~Monroe, \enquote{Quantum information processing with atoms and photons,}
  {\protect\JournalTitle{Nature}} \textbf{416}, 238--246 (2002).

\bibitem{Sasaki:11}
M.~Sasaki, M.~Fujiwara, H.~Ishizuka, W.~Klaus, K.~Wakui, M.~Takeoka, S.~Miki,
  T.~Yamashita, Z.~Wang, A.~Tanaka, K.~Yoshino, Y.~Nambu, S.~Takahashi,
  A.~Tajima, A.~Tomita, T.~Domeki, T.~Hasegawa, Y.~Sakai, H.~Kobayashi,
  T.~Asai, K.~Shimizu, T.~Tokura, T.~Tsurumaru, M.~Matsui, T.~Honjo, K.~Tamaki,
  H.~Takesue, Y.~Tokura, J.~F. Dynes, A.~R. Dixon, A.~W. Sharpe, Z.~L. Yuan,
  A.~J. Shields, S.~Uchikoga, M.~Legr\'{e}, S.~Robyr, P.~Trinkler, L.~Monat,
  J.-B. Page, G.~Ribordy, A.~Poppe, A.~Allacher, O.~Maurhart, T.~L\"{a}nger,
  M.~Peev, and A.~Zeilinger, \enquote{Field test of quantum key distribution in
  the {Tokyo QKD} network,} {\protect\JournalTitle{Optics Express}}
  \textbf{19}, 10387--10409 (2011).

\bibitem{PhysRevLett.115.250401}
M.~Giustina, M.~A.~M. Versteegh, S.~Wengerowsky, J.~Handsteiner, A.~Hochrainer,
  K.~Phelan, F.~Steinlechner, J.~Kofler, J.-A. Larsson, C.~Abell\'an, W.~Amaya,
  V.~Pruneri, M.~W. Mitchell, J.~Beyer, T.~Gerrits, A.~E. Lita, L.~K. Shalm,
  S.~W. Nam, T.~Scheidl, R.~Ursin, B.~Wittmann, and A.~Zeilinger,
  \enquote{Significant-loophole-free test of bell's theorem with entangled
  photons,} {\protect\JournalTitle{Phys. Rev. Lett.}} \textbf{115}, 250401
  (2015).

\bibitem{dello2022advances}
S.~Dello~Russo, A.~Elefante, D.~Dequal, D.~K. Pallotti, L.~Santamaria~Amato,
  F.~Sgobba, and M.~Siciliani~de Cumis, \enquote{Advances in mid-infrared
  single-photon detection,} in \emph{Photonics,}  vol.~9 (MDPI, 2022), p. 470.

\bibitem{Taylor2019}
G.~G. Taylor, D.~Morozov, N.~R. Gemmell, K.~Erotokritou, S.~Miki, H.~Terai, and
  R.~H. Hadfield, \enquote{Photon counting {LIDAR} at 2.3 \textmu m wavelength
  with superconducting nanowires,} {\protect\JournalTitle{Opt. Express}}
  \textbf{27}, 38147--38158 (2019).

\bibitem{aellen2008feasibility}
T.~Aellen, M.~Giovannini, J.~Faist, and J.~P. Von~der Weid,
  \enquote{Feasibility study of free-space quantum key distribution in the
  mid-infrared,} {\protect\JournalTitle{Quantum Information and Computation}}
  \textbf{8}, 0001--0011 (2008).

\bibitem{Chen2017}
L.~Chen, D.~Schwarzer, V.~B. Verma, M.~J. Stevens, F.~Marsili, R.~P. Mirin,
  S.~W. Nam, and A.~M. Wodtke, \enquote{{Mid-infrared Laser-Induced
  Fluorescence with Nanosecond Time Resolution Using a Superconducting Nanowire
  Single-Photon Detector: New Technology for Molecular Science},}
  {\protect\JournalTitle{Accounts of Chemical Research}} \textbf{50},
  1400--1409 (2017).

\bibitem{wollman2021recent}
E.~E. Wollman, V.~B. Verma, A.~B. Walter, J.~Chiles, B.~Korzh, J.~P. Allmaras,
  Y.~Zhai, A.~E. Lita, A.~N. McCaughan, E.~Schmidt, S.~Frasca, R.~P. Mirin,
  S.-W. Nam, and M.~D. Shaw, \enquote{Recent advances in superconducting
  nanowire single-photon detector technology for exoplanet transit spectroscopy
  in the mid-infrared,} {\protect\JournalTitle{Journal of Astronomical
  Telescopes, Instruments, and Systems}} \textbf{7}, 011004 (2021).

\bibitem{anderson2022recent}
P.~D. Anderson, J.~D. Beck, W.~Sullivan~III, C.~Schaake, J.~McCurdy, M.~Skokan,
  P.~Mitra, and X.~Sun, \enquote{Recent advancements in {HgCdTe APDs} for space
  applications,} {\protect\JournalTitle{Journal of Electronic Materials}}
  \textbf{51}, 6803--6814 (2022).

\bibitem{gol2001picosecond}
G.~{Gol’tsman}, O.~Okunev, G.~Chulkova, A.~Lipatov, A.~Semenov, K.~Smirnov,
  B.~Voronov, A.~Dzardanov, C.~Williams, and R.~Sobolewski, \enquote{Picosecond
  superconducting single-photon optical detector,}
  {\protect\JournalTitle{Applied Physics Letters}} \textbf{79}, 705--707
  (2001).

\bibitem{Verma2021}
V.~B. Verma, B.~Korzh, A.~B. Walter, A.~E. Lita, R.~M. Briggs, M.~Colangelo,
  Y.~Zhai, E.~E. Wollman, A.~D. Beyer, J.~P. Allmaras, H.~Vora, D.~Zhu,
  E.~Schmidt, A.~G. Kozorezov, K.~K. Berggren, R.~P. Mirin, S.~W. Nam, and
  M.~D. Shaw, \enquote{Single-photon detection in the mid-infrared up to 10
  \textmu m wavelength using tungsten silicide superconducting nanowire
  detectors,} {\protect\JournalTitle{APL Photonics}} \textbf{6}, 056101 (2021).

\bibitem{morozov2022superconducting}
D.~V. Morozov, A.~Casaburi, and R.~H. Hadfield, \enquote{Superconducting photon
  detectors,} {\protect\JournalTitle{Contemporary Physics}} pp. 1--23 (2022).

\bibitem{seager2008exoplanet}
S.~Seager, \enquote{Exoplanet transit spectroscopy and photometry,}
  {\protect\JournalTitle{Space Science Reviews}} \textbf{135}, 345--354 (2008).

\bibitem{boccaletti2005imaging}
A.~Boccaletti, P.~Baudoz, J.~Baudrand, J.~Reess, and D.~Rouan, \enquote{Imaging
  exoplanets with the coronagraph of {JWST/MIRI},}
  {\protect\JournalTitle{Advances in Space Research}} \textbf{36}, 1099--1106
  (2005).

\bibitem{kaltenegger2019characterize}
L.~Kaltenegger, \enquote{How to characterize habitable worlds and signs of
  life,} {\protect\JournalTitle{arXiv preprint arXiv:1911.05597}}  (2019).

\bibitem{martin2012high}
S.~Martin, A.~Booth, K.~Liewer, N.~Raouf, F.~Loya, and H.~Tang, \enquote{High
  performance testbed for four-beam infrared interferometric nulling and
  exoplanet detection,} {\protect\JournalTitle{Applied Optics}} \textbf{51},
  3907--3921 (2012).

\bibitem{konrad2021large}
B.~Konrad, E.~Alei, D.~Angerhausen, {\'O}.~Carri{\'o}n-Gonz{\'a}lez, J.~J.
  Fortney, J.~L. Grenfell, D.~Kitzmann, P.~Molli{\`e}re, S.~Rugheimer,
  F.~Wunderlich, and S.~P. Quanz, \enquote{Large interferometer for exoplanets
  {(LIFE)}: {III}. spectral resolution, wavelength range and sensitivity
  requirements based on atmospheric retrieval analyses of an exo-earth,}
  {\protect\JournalTitle{arXiv preprint arXiv:2112.02054}}  (2021).

\bibitem{quanz2018exoplanet}
S.~P. Quanz, J.~Kammerer, D.~Defr{\`e}re, O.~Absil, A.~M. Glauser, and
  D.~Kitzmann, \enquote{Exoplanet science with a space-based mid-infrared
  nulling interferometer,} in \emph{Optical and Infrared Interferometry and
  Imaging VI,}  vol. 10701 (SPIE, 2018), pp. 415--431.

\bibitem{peeters2002iso}
E.~Peeters, N.~Mart{\'\i}n-Hern{\'a}ndez, F.~Damour, P.~Cox, P.~Roelfsema,
  J.-P. Baluteau, A.~Tielens, E.~Churchwell, M.~Kessler, J.~Mathis,
  C.~Morriset, and D.~Schaerer, \enquote{{ISO} spectroscopy of compact {HII}
  regions in the {Galaxy-I}. {The} catalogue,} {\protect\JournalTitle{Astronomy
  \& Astrophysics}} \textbf{381}, 571--605 (2002).

\bibitem{palotas2020infrared}
J.~Palot{\'a}s, J.~Martens, G.~Berden, and J.~Oomens, \enquote{The infrared
  spectrum of protonated buckminsterfullerene {C60H+},}
  {\protect\JournalTitle{Nature Astronomy}} \textbf{4}, 240--245 (2020).

\bibitem{golwala2022novel}
S.~R. Golwala and E.~Figueroa-Feliciano, \enquote{Novel quantum sensors for
  light dark matter and neutrino detection,} {\protect\JournalTitle{Annual
  Review of Nuclear and Particle Science}} \textbf{72}, 419--446 (2022).

\bibitem{PhysRevLett.128.131801}
J.~Liu, K.~Dona, G.~Hoshino, S.~Knirck, N.~Kurinsky, M.~Malaker, D.~W. Miller,
  A.~Sonnenschein, M.~H. Awida, P.~S. Barry, K.~K. Berggren, D.~Bowring,
  G.~Carosi, C.~Chang, A.~Chou, R.~Khatiwada, S.~Lewis, J.~Li, S.~W. Nam,
  O.~Noroozian, and T.~X. Zhou, \enquote{Broadband solenoidal haloscope for
  terahertz axion detection,} {\protect\JournalTitle{Phys. Rev. Lett.}}
  \textbf{128}, 131801 (2022).

\bibitem{chiles2022new}
J.~Chiles, I.~Charaev, R.~Lasenby, M.~Baryakhtar, J.~Huang, A.~Roshko,
  G.~Burton, M.~Colangelo, K.~Van~Tilburg, A.~Arvanitaki, S.~W. Nam, and K.~K.
  Berggren, \enquote{New constraints on dark photon dark matter with
  superconducting nanowire detectors in an optical haloscope,}
  {\protect\JournalTitle{Physical Review Letters}} \textbf{128}, 231802 (2022).

\bibitem{hochberg2022new}
Y.~Hochberg, B.~V. Lehmann, I.~Charaev, J.~Chiles, M.~Colangelo, S.~W. Nam, and
  K.~K. Berggren, \enquote{New constraints on dark matter from superconducting
  nanowires,} {\protect\JournalTitle{Physical Review D}} \textbf{106}, 112005
  (2022).

\bibitem{lau2023superconducting}
J.~A. Lau, V.~B. Verma, D.~Schwarzer, and A.~M. Wodtke,
  \enquote{Superconducting single-photon detectors in the mid-infrared for
  physical chemistry and spectroscopy,} {\protect\JournalTitle{Chemical Society
  Reviews}}  (2023).

\bibitem{ressler2008performance}
M.~E. Ressler, H.~Cho, R.~A. Lee, K.~G. Sukhatme, J.~J. Drab, G.~Domingo, M.~E.
  McKelvey, R.~E. McMurray~Jr, and J.~L. Dotson, \enquote{Performance of the
  {JWST/MIRI} {Si:As} detectors,} in \emph{High Energy, Optical, and Infrared
  Detectors for Astronomy III,}  vol. 7021 (SPIE, 2008), pp. 224--235.

\bibitem{wada2010mid}
T.~Wada and H.~Kataza, \enquote{Mid-infrared camera without lens {(MIRACLE) for
  SPICA},} in \emph{Space Telescopes and Instrumentation 2010: Optical,
  Infrared, and Millimeter Wave,}  vol. 7731 (SPIE, 2010), pp. 272--282.

\bibitem{ressler2015mid}
M.~Ressler, K.~Sukhatme, B.~Franklin, J.~Mahoney, M.~Thelen, P.~Bouchet,
  J.~Colbert, M.~Cracraft, D.~Dicken, R.~Gastaud, G.~Goodson, P.~Eccleston,
  V.~Moreau, G.~Rieke, and A.~Schneider, \enquote{The mid-infrared instrument
  for the {James Webb Space Telescope, VIII: the MIRI focal plane system},}
  {\protect\JournalTitle{Publications of the Astronomical Society of the
  Pacific}} \textbf{127}, 675 (2015).

\bibitem{greene2016characterizing}
T.~P. Greene, M.~R. Line, C.~Montero, J.~J. Fortney, J.~Lustig-Yaeger, and
  K.~Luther, \enquote{Characterizing transiting exoplanet atmospheres with
  {JWST},} {\protect\JournalTitle{The Astrophysical Journal}} \textbf{817}, 17
  (2016).

\bibitem{cabrera2019characterization}
M.~S. Cabrera, C.~W. McMurtry, W.~J. Forrest, J.~L. Pipher, M.~L. Dorn, and
  D.~L. Lee, \enquote{Characterization of a 15-\textmu m cutoff {HgCdTe}
  detector array for astronomy,} {\protect\JournalTitle{Journal of Astronomical
  Telescopes, Instruments, and Systems}} \textbf{6}, 011004 (2019).

\bibitem{pipher2021mid}
J.~L. Pipher, C.~W. McMurtry, M.~S. Cabrera, and W.~J. Forrest,
  \enquote{Mid-infrared detector array technologies for {SOFIA} and sub-orbital
  observatory instruments,} {\protect\JournalTitle{Journal of Astronomical
  Instrumentation}} \textbf{10}, 2150008 (2021).

\bibitem{rogalski2019type}
A.~Rogalski, P.~Martyniuk, and M.~Kopytko, \enquote{Type-{II} superlattice
  photodetectors versus {HgCdTe} photodiodes,} {\protect\JournalTitle{Progress
  in Quantum Electronics}} \textbf{68}, 100228 (2019).

\bibitem{roellig2020mid}
T.~L. Roellig, C.~W. McMurtry, T.~P. Greene, T.~Matsuo, I.~Sakon, and J.~G.
  Staguhn, \enquote{Mid-infrared detector development for the {Origins Space
  Telescope},} {\protect\JournalTitle{Journal of Astronomical Telescopes,
  Instruments, and Systems}} \textbf{6}, 041503 (2020).

\bibitem{hailey2021kinetic}
S.~Hailey-Dunsheath, R.~M. Janssen, J.~Glenn, C.~M. Bradford, J.~Perido,
  J.~Redford, and J.~Zmuidzinas, \enquote{Kinetic inductance detectors for the
  {Origins} space telescope,} {\protect\JournalTitle{Journal of Astronomical
  Telescopes, Instruments, and Systems}} \textbf{7}, 011015 (2021).

\bibitem{perido2020extending}
J.~Perido, J.~Glenn, P.~Day, A.~Fyhrie, H.~Leduc, J.~Zmuidzinas, and
  C.~McKenney, \enquote{Extending {KIDs} to the mid-{IR} for future space and
  suborbital observatories,} {\protect\JournalTitle{Journal of Low Temperature
  Physics}} \textbf{199}, 696--703 (2020).

\bibitem{lamas2013nanosecond}
A.~Lamas-Linares, B.~Calkins, N.~A. Tomlin, T.~Gerrits, A.~E. Lita, J.~Beyer,
  R.~P. Mirin, and S.~Woo~Nam, \enquote{Nanosecond-scale timing jitter for
  single photon detection in transition edge sensors,}
  {\protect\JournalTitle{Applied Physics Letters}} \textbf{102}, 231117 (2013).

\bibitem{Reddy2020}
D.~V. Reddy, R.~R. Nerem, S.~W. Nam, R.~P. Mirin, and V.~B. Verma,
  \enquote{Superconducting nanowire single-photon detectors with 98\% system
  detection efficiency at 1550 nm,} {\protect\JournalTitle{Optica}} \textbf{7},
  1649--1653 (2020).

\bibitem{hu2020detecting}
P.~Hu, H.~Li, L.~You, H.~Wang, Y.~Xiao, J.~Huang, X.~Yang, W.~Zhang, Z.~Wang,
  and X.~Xie, \enquote{Detecting single infrared photons toward optimal system
  detection efficiency,} {\protect\JournalTitle{Optics Express}} \textbf{28},
  36884--36891 (2020).

\bibitem{chang2021detecting}
J.~Chang, J.~Los, J.~Tenorio-Pearl, N.~Noordzij, R.~Gourgues, A.~Guardiani,
  J.~Zichi, S.~Pereira, H.~Urbach, V.~Zwiller, S.~Dorenbos, and Z.~IE,
  \enquote{Detecting telecom single photons with 99.5- 2.07+ 0.5\% system
  detection efficiency and high time resolution,} {\protect\JournalTitle{APL
  Photonics}} \textbf{6}, 036114 (2021).

\bibitem{korzh2020demonstration}
B.~Korzh, Q.~Y. Zhao, J.~P. Allmaras, S.~Frasca, T.~M. Autry, E.~A. Bersin,
  A.~D. Beyer, R.~M. Briggs, B.~Bumble, M.~Colangelo, G.~M. Crouch, A.~E. Dane,
  T.~Gerrits, A.~E. Lita, F.~Marsili, G.~Moody, C.~Pe{\~{n}}a, E.~Ramirez,
  J.~D. Rezac, N.~Sinclair, M.~J. Stevens, A.~E. Velasco, V.~B. Verma, E.~E.
  Wollman, S.~Xie, D.~Zhu, P.~D. Hale, M.~Spiropulu, K.~L. Silverman, R.~P.
  Mirin, S.~W. Nam, A.~G. Kozorezov, M.~D. Shaw, and K.~K. Berggren,
  \enquote{{Demonstration of sub-3 ps temporal resolution with a
  superconducting nanowire single-photon detector},}
  {\protect\JournalTitle{Nat. Photonics}} \textbf{14}, 250--255 (2020).

\bibitem{chiles2020superconducting}
J.~Chiles, S.~M. Buckley, A.~Lita, V.~B. Verma, J.~Allmaras, B.~Korzh, M.~D.
  Shaw, J.~M. Shainline, R.~P. Mirin, and S.~W. Nam, \enquote{Superconducting
  microwire detectors with single-photon sensitivity in the near-infrared,}
  {\protect\JournalTitle{arXiv e-prints}} p.~{\tt arXiv:2002.12858} (2020).

\bibitem{wang2022twin}
S.~Wang, Z.-Q. Yin, D.-Y. He, W.~Chen, R.-Q. Wang, P.~Ye, Y.~Zhou, G.-J.
  Fan-Yuan, F.-X. Wang, Y.-G. Zhu, P.~V. Morozov, A.~V. Divochiy, Z.~Zhou,
  G.-C. Guo, and Z.-F. Han, \enquote{Twin-field quantum key distribution over
  830-km fibre,} {\protect\JournalTitle{Nature Photonics}} \textbf{16},
  154--161 (2022).

\bibitem{craiciu2023high}
I.~Craiciu, B.~Korzh, A.~D. Beyer, A.~Mueller, J.~P. Allmaras, L.~Narv{\'a}ez,
  M.~Spiropulu, B.~Bumble, T.~Lehner, E.~E. Wollman, and M.~D. Shaw,
  \enquote{High-speed detection of 1550 nm single photons with superconducting
  nanowire detectors,} {\protect\JournalTitle{Optica}} \textbf{10}, 183--190
  (2023).

\bibitem{Marsili2012a}
F.~Marsili, F.~Bellei, F.~Najafi, A.~E. Dane, E.~A. Dauler, R.~J. Molnar, and
  K.~K. Berggren, \enquote{{Efficient Single Photon Detection from 500 nm to 5
  \textmu m Wavelength},} {\protect\JournalTitle{Nano Letters}} \textbf{12},
  4799--4804 (2012).

\bibitem{colangelo2022large}
M.~Colangelo, A.~B. Walter, B.~A. Korzh, E.~Schmidt, B.~Bumble, A.~E. Lita,
  A.~D. Beyer, J.~P. Allmaras, R.~M. Briggs, A.~G. Kozorezov, E.~E. Wollman,
  M.~D. Shaw, and K.~K. Berggren, \enquote{Large-area superconducting nanowire
  single-photon detectors for operation at wavelengths up to 7.4 \textmu m,}
  {\protect\JournalTitle{Nano Letters}}  (2022).

\bibitem{korneev2012nbn}
A.~Korneev, Y.~Korneeva, I.~Florya, B.~Voronov, and G.~Goltsman, \enquote{{NbN}
  nanowire superconducting single-photon detector for mid-infrared,}
  {\protect\JournalTitle{Physics Procedia}} \textbf{36}, 72--76 (2012).

\bibitem{pan2022mid}
Y.~Pan, H.~Zhou, X.~Zhang, H.~Yu, L.~Zhang, M.~Si, H.~Li, L.~You, and Z.~Wang,
  \enquote{Mid-infrared {Nb4N3-based} superconducting nanowire single photon
  detectors for wavelengths up to 10 $\mu$m,} {\protect\JournalTitle{Optics
  Express}} \textbf{30}, 40044--40052 (2022).

\bibitem{chen2020mid}
Q.~Chen, R.~Ge, L.~Zhang, F.~Li, B.~Zhang, Y.~Dai, Y.~Fei, X.~Wang, X.~Jia,
  Q.~Zhao, X.~Tu, L.~Kang, J.~Chen, and P.~Wu, \enquote{Mid-infrared single
  photon detector with superconductor {Mo}80{Si}20 nanowire,}
  {\protect\JournalTitle{arXiv preprint arXiv:2011.06699}}  (2020).

\bibitem{chang2022efficient}
J.~Chang, J.~W. Los, R.~Gourgues, S.~Steinhauer, S.~Dorenbos, S.~F. Pereira,
  H.~P. Urbach, V.~Zwiller, and I.~E. Zadeh, \enquote{Efficient mid-infrared
  single-photon detection using superconducting {NbTiN} nanowires with high
  time resolution in a {Gifford-McMahon} cryocooler,}
  {\protect\JournalTitle{Photonics Research}} \textbf{10}, 1063--1070 (2022).

\bibitem{vodolazov2017single}
D.~Y. Vodolazov, \enquote{Single-photon detection by a dirty current-carrying
  superconducting strip based on the kinetic-equation approach,}
  {\protect\JournalTitle{Physical Review Applied}} \textbf{7}, 034014 (2017).

\bibitem{Frasca2019}
S.~Frasca, B.~Korzh, M.~Colangelo, D.~Zhu, A.~E. Lita, J.~P. Allmaras, E.~E.
  Wollman, V.~B. Verma, A.~E. Dane, E.~Ramirez, A.~D. Beyer, S.~W. Nam, A.~G.
  Kozorezov, M.~D. Shaw, and K.~K. Berggren, \enquote{{Determining the
  depairing current in superconducting nanowire single-photon detectors},}
  {\protect\JournalTitle{Phys. Rev. B}} \textbf{100}, 054520 (2019).

\bibitem{annunziata2010reset}
A.~J. Annunziata, O.~Quaranta, D.~F. Santavicca, A.~Casaburi, L.~Frunzio,
  M.~Ejrnaes, M.~J. Rooks, R.~Cristiano, S.~Pagano, A.~Frydman, and D.~E.
  Prober, \enquote{Reset dynamics and latching in niobium superconducting
  nanowire single-photon detectors,} {\protect\JournalTitle{Journal of Applied
  Physics}} \textbf{108}, 084507 (2010).

\bibitem{Walter2023}
A.~B. Walter, E.~Schmidt, M.~Colangelo, I.~Craiciu, G.~G. Taylor, R.~M. Briggs,
  J.~P. Allmaras, B.~Bumble, A.~D. Beyer, E.~E. Wollman, K.~K. Berggren, M.~D.
  Shaw, and B.~Korzh, \enquote{{Cryogenic operation of broadband thermal
  infrared emitters},} {\protect\JournalTitle{arXiv}}  (2023).

\bibitem{doi:10.1063/1.4881981}
L.~Zhang, L.~You, D.~Liu, W.~Zhang, L.~Zhang, X.~Liu, J.~Wu, Y.~He, C.~Lv,
  Z.~Wang, and X.~Xie, \enquote{Characterization of superconducting nanowire
  single-photon detector with artificial constrictions,}
  {\protect\JournalTitle{AIP Advances}} \textbf{4}, 067114 (2014).

\bibitem{du2022sources}
P.~Du, D.~Egana-Ugrinovic, R.~Essig, and M.~Sholapurkar, \enquote{Sources of
  low-energy events in low-threshold dark-matter and neutrino detectors,}
  {\protect\JournalTitle{Physical Review X}} \textbf{12}, 011009 (2022).

\bibitem{patil2012comparison}
V.~Patil and H.~Kulkarni, \enquote{Comparison of confidence intervals for the
  poisson mean: some new aspects,} {\protect\JournalTitle{REVSTAT-Statistical
  Journal}} \textbf{10}, 211--22 (2012).

\bibitem{baek2009superconducting}
B.~Baek, J.~A. Stern, and S.~W. Nam, \enquote{Superconducting nanowire
  single-photon detector in an optical cavity for front-side illumination,}
  {\protect\JournalTitle{Applied Physics Letters}} \textbf{95}, 191110 (2009).

\bibitem{heath2015nanoantenna}
R.~M. Heath, M.~G. Tanner, T.~D. Drysdale, S.~Miki, V.~Giannini, S.~A. Maier,
  and R.~H. Hadfield, \enquote{Nanoantenna enhancement for telecom-wavelength
  superconducting single photon detectors,} {\protect\JournalTitle{Nano
  Letters}} \textbf{15}, 819--822 (2015).

\bibitem{xu2021superconducting}
Y.~Xu, A.~Kuzmin, E.~Knehr, M.~Blaicher, K.~Ilin, P.-I. Dietrich, W.~Freude,
  M.~Siegel, and C.~Koos, \enquote{Superconducting nanowire single-photon
  detector with {3D}-printed free-form microlenses,}
  {\protect\JournalTitle{Optics Express}} \textbf{29}, 27708--27731 (2021).

\bibitem{hadfield2009single}
R.~H. Hadfield, \enquote{Single-photon detectors for optical quantum
  information applications,} {\protect\JournalTitle{Nature Photonics}}
  \textbf{3}, 696--705 (2009).

\bibitem{baselmans2022ultra}
J.~Baselmans, F.~Facchin, A.~P. Laguna, J.~Bueno, D.~Thoen, V.~Murugesan,
  N.~Llombart, and P.~De~Visser, \enquote{Ultra-sensitive {THz} microwave
  kinetic inductance detectors for future space telescopes,}
  {\protect\JournalTitle{Astronomy \& Astrophysics}} \textbf{665}, A17 (2022).

\bibitem{zhu2019superconducting}
D.~Zhu, M.~Colangelo, B.~A. Korzh, Q.-Y. Zhao, S.~Frasca, A.~E. Dane, A.~E.
  Velasco, A.~D. Beyer, J.~P. Allmaras, E.~Ramirez, W.~J. Strickland, D.~F.
  Santavicca, M.~D. Shaw, and K.~K. Berggren, \enquote{Superconducting nanowire
  single-photon detector with integrated impedance-matching taper,}
  {\protect\JournalTitle{Applied Physics Letters}} \textbf{114}, 042601 (2019).

\bibitem{marsili2011single}
F.~Marsili, F.~Najafi, E.~Dauler, F.~Bellei, X.~Hu, M.~Csete, R.~J. Molnar, and
  K.~K. Berggren, \enquote{Single-photon detectors based on ultranarrow
  superconducting nanowires,} {\protect\JournalTitle{Nano Letters}}
  \textbf{11}, 2048--2053 (2011).

\end{thebibliography}

\end{document}